\documentstyle[12pt,epsfig]{article}

\def\gev{{\rm Ge}\kern-1.pt{\rm V}}
\def\3{\ss}

\mathchardef\less=316
\mathchardef\greater=318

\begin{document}

\newcommand{\ETCONE}{$E_{\perp}^{\theta > 10^{\circ}}$}
\newcommand{\YJB}{$y_{\mathsf{JB}}$}
 
\def\gp{$ {\gamma}p $}
\def\W{$ W $}
\def\sgpccx{ $ {\sigma}_{{\gamma}p {\rightarrow} c\overline{c}X} $ }
\def\Ds{ $ D^{\ast}(2010)^{\pm} $ }
\def\ds{D^{\ast}}
\def\d0{D^{0}}
\def\dspm{{\ds}^{\pm}}
\def\dskpi{ {\ds}^{+} \rightarrow \d0 \pi^{+}_{S}%
        \rightarrow (K^{-} \pi^{+}) \pi^{+}_{S} }
\def\dsk3pi{ {\ds}^{+} \rightarrow \d0 \pi^{+}_{S}%
        \rightarrow (K^{-} \pi^{+} \pi^{+} \pi^{-}) \pi^{+}_{S} }
\def\lumi{ 2.99\,pb$^{-1}$}
\def\qq{$Q^{2}$}
\def\g2{GeV$^{2}$}
\def\Decay{ D^{\ast + }~\rightarrow~(D^{o}~%
    \rightarrow~K^{-}\pi^{+})~\pi^{+}~+~(c.c.)}
\def\Dec{ D^{\ast + }~\rightarrow~(D^{o}~%
        \rightarrow~K^{-}\pi^{+})~\pi^{+}}
\def\bethe{$e^{+}p \rightarrow e^{+}{\gamma}p$}
\def\wrang{$115 < W < 280$\,GeV}
\def\ptrang{$p_{\perp}^{\ds} > 3$\,GeV}
\def\etarang{$-1.5 < \eta^{\ds} < 1.0 $}
\def\qqrang{\qq~$<$~4\,\g2}
\def\totala{$152 \pm 16 $}
\def\totalb{$167 \pm25$}
\def\xsec{$(10.6 \pm 1.7 \pm^{1.6}_{1.3})$\,nb}
\def\xsecs{$(10.6 \pm 1.7 ({\it stat.})\pm^{1.6}_{1.3}%
            ({\it syst.}))$\,nb}
  
 
\vspace{7.0cm}
 
\title { \bf 
    { Differential Cross Sections of \boldmath ${ \ds}^\pm$ Photoproduction
     in $ep$~Collisions at HERA }
     \author {ZEUS Collaboration } }
\date{ } 
\maketitle

\begin{abstract}
Inclusive photoproduction of $\dspm$ in $ep$ collisions at HERA    
has been measured with the ZEUS detector for photon-proton 
centre of mass energies in the range \linebreak
\wrang\ and 
photon virtuality $Q^2 <$ 4\,\g2.
The cross section
$\sigma_{ep\,\rightarrow\,\ds X}\,$                                  
integrated over                       
the kinematic region \ptrang~and \etarang\
is {\xsecs}.
Differential cross sections as functions of $p_{\perp}^{\ds}$, $\eta^{\ds}$ and $W$ 
are given. The data are compared with two next-to-leading order 
perturbative QCD predictions. For a calculation using a massive charm scheme            
the predicted cross sections are smaller than the measured ones. 
A recent calculation using a massless charm scheme is in 
agreement with the data.
\end{abstract}

\vspace{-12.5 cm}
\hspace{12 cm}
{DESY-97-026}

\setcounter{page}{0} 
\thispagestyle{empty} 
\vspace{3cm}

\newpage
 
\begin{center}                                                         
  {\Large  The ZEUS Collaboration }                               
\end{center}
{ \small 
  \parindent 0.pt                                                                          
  J.~Breitweg,
  M.~Derrick,
  D.~Krakauer,
  S.~Magill,
  D.~Mikunas, 
  B.~Musgrave,
  J.~Repond,
  R.~Stanek,
  R.L.~Talaga, 
  R.~Yoshida, 
  H.~Zhang  \\                                                                                     
 {\it Argonne National Laboratory, Argonne, IL, USA}~$^{p}$
% ...

\newpage 

% ...
$^{  21}$ now at ATLAS Collaboration, Univ. of Munich \\ 
% ...

\begin{tabular}[h]{rp{14cm}} 
$^{a}$ &  supported by the Natural Sciences and Engineering Research Council of Canada (NSERC)  \\ 
$^{b}$ &  supported by the FCAR of Qu\'ebec, Canada  \\ 
$^{c}$ &  supported by the German Federal Ministry for Education and 
          Science, Research and Technology (BMBF), under contract   
          numbers 057BN19P, 057FR19P, 057HH19P, 057HH29P, 057SI75I \\ 
$^{d}$ &  supported by the MINERVA Gesellschaft f\"ur Forschung GmbH,  
          the German Israeli Foundation, and the U.S.-Israel Binational  
          Science Foundation \\ 
$^{e}$ &  supported by the German Israeli Foundation, and 
          by the Israel Academy of Science  \\                                                     
$^{f}$ &  supported by the Italian National Institute for Nuclear Physics                          
          (INFN) \\                                                                                
$^{g}$ &  supported by the Japanese Ministry of Education, Science and                             
          Culture (the Monbusho) and its grants for Scientific Research \\                         
$^{h}$ &  supported by the Korean Ministry of Education and Korea Science                          
          and Engineering Foundation  \\                                                           
$^{i}$ &  supported by the Netherlands Foundation for Research on                                  
          Matter (FOM) \\                                                                          
$^{j}$ &  supported by the Polish State Committee for Scientific                                   
          Research, grant No.~115/E-343/SPUB/P03/120/96  \\                                        
$^{k}$ &  supported by the Polish State Committee for Scientific                                   
          Research (grant No. 2 P03B 083 08) and Foundation for                                    
          Polish-German Collaboration  \\                                                          
$^{l}$ &  partially supported by the German Federal Ministry for                                   
          Education and Science, Research and Technology (BMBF)  \\                                
$^{m}$ &  supported by the German Federal Ministry for Education and                               
          Science, Research and Technology (BMBF), and the Fund of                                 
          Fundamental Research of Russian Ministry of Science and                                  
          Education and by INTAS-Grant No. 93-63 \\                                                
$^{n}$ &  supported by the Spanish Ministry of Education                                           
          and Science through funds provided by CICYT \\                                           
$^{o}$ &  supported by the Particle Physics and                                                    
          Astronomy Research Council \\                                                            
$^{p}$ &  supported by the US Department of Energy \\                                              
$^{q}$ &  supported by the US National Science Foundation \\ 

\end{tabular}

  } 
   
\newpage
\section{ \bf Introduction }
\pagenumbering{arabic}
\setcounter{page}{1} 
 
Production of the heavy quarks $c$ and $b$ at HERA
is dominated by photoproduction, where a 
quasi-real photon with negative
four momentum squared, $Q^2$, 
close to zero is emitted by the incoming electron and
interacts with the proton.                     
Heavy quark photoproduction
can be used to probe
perturbative QCD calculations
with a hard scale stemming from
the heavy quark mass and the high transverse momentum          
of the produced parton.                                   
At leading order (LO) in QCD
two types of processes are responsible for the production 
of heavy quarks:                       
the direct photon processes,
where the photon participates as a point-like particle which
interacts with a parton from the incoming proton, 
and the resolved photon processes, where 
the photon is a source of partons, one of which         
scatters off a parton from the proton.
Charm quarks present in the parton distributions of the photon, as
well as of the proton, lead to processes
like $cg\to cg$, which are called charm flavour excitation.     
In next-to-leading order (NLO) QCD only the sum of direct and resolved
processes is unambiguously defined. 
Two types of NLO calculations using different approaches are available
for comparison with measurements of charm photoproduction at HERA.
The massive charm approach~\cite{Ellis,NLOtot, NLOdiff} assumes light quarks to be 
the only active flavours within the structure functions of
the proton and the photon, while the massless charm
approach~\cite{kkks,cacciari} 
also treats charm as an active flavour.
 
The total charm photoproduction cross section 
\sgpccx has recently been measured at HERA~\cite{paper_93,H1_94}
at a \gp~centre of mass energy $W \approx 200$\,GeV, and was found
to be about one order of magnitude larger than fixed
target data. These measurements were compared with NLO QCD                  
calculations~\cite{NLOtot}
and with a calculation based on a semihard approach of    
QCD~\cite{zotov}.
The precision of the comparison was
limited by large systematic uncertainties 
in both data and theory.
In the case of the data the uncertainty was due to the
necessity of extrapolating the measured cross sections into kinematic regions
not accessible to the experiments. 
The uncertainties of the theoretical predictions were
generated by the dependence of the NLO calculations on      
the charm mass, $m_c$,         
the QCD renormalization and factorization scales,
and on the parton density parametrizations assumed
for the proton and the photon.

In this study we use a sample of $D^{\ast}(2010)^\pm$ mesons
collected with the ZEUS detector during 1994.
The sixfold increase of the data sample with respect to our previous
analysis~\cite{paper_93}  allowed the measurement
of the differential distributions in $p_\perp^{\ds}$, $\eta^{\ds}$
and $W$, where $p_{\perp}^{\ds}$ is the $\ds$
transverse momentum with respect to the beam 
axis and $\eta^{\ds}$ is the pseudorapidity%
\footnote{The pseudorapidity $\eta$ is defined as $-\ln(\tan \frac{\theta}{2})$,
where the polar angle $\theta$ is taken with respect to the proton beam direction.}
of the $\ds$. The measurement of the $ep \rightarrow D^*X$ cross sections 
was performed in the restricted kinematic
region $Q^2 < 4$\,GeV$^2$, $115 < W < 280$\,GeV, \ptrang\ and \etarang.
The results are compared with NLO QCD predictions calculated in both
the massive and the massless charm approach.
The theoretical uncertainties in the calculations are strongly reduced 
in the restricted kinematic region, thus allowing
a more precise comparison of the perturbative QCD calculations
with our data. 
The results of the calculations using a massless charm scheme are
sensitive to the charm content of the photon and are insensitive to that
of the proton \cite{kkks,kniehl}.
 
$\ds$ mesons are                             
reconstructed from their decay products through the two decay 
modes\footnote{In this analysis the charge conjugated 
processes are also included.}:
\begin{equation}
   \dskpi,
   \label{kpi}
\end{equation}
\begin{equation}
   \dsk3pi.
   \label{k3pi}
\end{equation}

\noindent
The small mass difference
$ M(\ds) - M(\d0) =  145.42 \pm 0.05 $\,MeV~\cite{PDG} yields a low momentum
pion (``soft pion'', $\pi_{S}$) from the $D^*$ decay and
prominent signals just above the threshold of the 
$ M(K\pi\pi_{S}) - M(K\pi) $ and
$M(K\pi\pi\pi\pi_{S}) - M(K\pi\pi\pi) $ distributions, where the phase space 
contribution is highly            
suppressed~\cite{DELTAM}.

\section{Experimental Conditions}
 
The data presented in this analysis were collected during 
the 1994 running period using the ZEUS detector 
at HERA, where a positron beam with energy
$E_{e} =$ 27.5\,GeV collided with a proton beam with energy
$E_{p} =$ 820\,GeV.             
A total of 153 colliding 
bunches were stored in 
HERA, together with 
additional 17~proton and 15~positron unpaired bunches  intended for 
studies of beam-induced backgrounds.
The time interval between bunch crossings was 96\,ns, and the
typical instantaneous luminosity was
1.5{$\cdot$}10$^{30}$\,cm$^{-2}$s$^{-1}$.
The r.m.s. of the
vertex position 
distribution along the beam direction was 12\,cm. 
The total integrated luminosity 
used in this analysis is \lumi. 
 
\subsection{The ZEUS Detector}
 
A detailed description of the ZEUS detector 
can be found in ref.~\cite{ZEUS1,ZEUS2}.
Here we present a brief
description of the components relevant to the present analysis. 
 
Charged particles are measured by the Central Tracking Detector 
(CTD)~\cite{CTD} which operates in a magnetic 
field of 1.43\,T provided by a thin 
superconducting solenoid. The CTD is a drift chamber consisting of 
72~cylindrical layers, arranged in 9 superlayers. Superlayers with  
wires parallel to the beam axis alternate with those inclined at a small 
angle to give a stereo view. The single hit efficiency of the CTD is greater than 
95$\%$ and the measured resolution in transverse 
momentum for tracks with hits in all the superlayers is 
$ \sigma_{p_{\perp}}/p_{\perp} = 0.005 p_{\perp} \bigoplus 0.016 $ ($p_{\perp}$ in~GeV).

Surrounding the solenoid is the uranium-scintillator calorimeter (CAL)~\cite{CAL}, 
which is divided into three parts: 
forward,%
\footnote{Throughout
this paper we use the standard ZEUS right-handed coordinate system, in which
$X = Y = Z = 0$ is the nominal interaction point, the positive
$Z$-axis points in the direction           of the protons (referred
to as the forward direction) and the $X$-axis is horizontal, pointing towards
the centre of HERA.}
barrel and rear covering the polar regions
$2.6^\circ$ to $36.7^\circ$,
$36.7^\circ$ to $129.1^\circ$ and
$129.1^\circ$ to $176.2^\circ$, respectively. 
The CAL covers 99.7$\%$ of the solid angle, with 
holes of $ 20 \times 20 $ cm$^{2}$ in the centres of                       
the forward and rear calorimeters to 
accommodate the HERA beam pipe. Each of the calorimeter parts is subdivided
into towers which         are segmented longitudinally into electromagnetic (EMC) 
and hadronic (HAC) sections. These sections are further subdivided into cells
each of which is read out by two photomultipliers. From test beam data, energy resolutions 
of ${\sigma}_E/E = 0.18/\sqrt{E}$ for electrons and 
${\sigma}_E/E = 0.35/\sqrt{E}$ for hadrons ($E$ in~GeV) have been obtained.
The timing resolution of a calorimeter cell is less than 1\,ns for energy deposits
greater than 4.5\,GeV. In order to minimise the effects of noise due to the uranium
radioactivity on the measurements, all EMC\,(HAC) cells with an energy deposit of less
than 60\,(110)\,MeV are discarded from the analysis. For cells without energy
deposits in neighbouring cells this cut was increased to 80\,(120)\,MeV.

Proton-gas events occurring in front of the nominal $ep$ interaction region are out
of time with respect to the $ep$ interactions and may thus be rejected by timing 
measurements made by the CAL and by scintillation veto counter arrays located
at $Z = -730$\,cm, $Z = -315$\,cm and $Z = -150$\,cm.
  
The luminosity was determined from the rate of the Bethe-Heitler process \bethe,
where the photon is measured by a calorimeter~\cite{LUMI} located at $Z = -107$\,m
in the HERA tunnel in the direction of the positron beam.

\subsection{Trigger}
 
The ZEUS detector uses a three level trigger system~\cite{ZEUS1}.
For the analysis presented in this paper the following trigger strategy was 
chosen to identify $\ds$ candidates
in the central region of the detector. 
 
In the first level trigger       the calorimeter cells were combined to define 
regional and global sums which were required to exceed one 
of the following thresholds:
\begin{itemize}
\item{} Total energy greater than 15\,GeV, 
\item{} Total EMC energy greater than 10\,GeV, 
\item{} Total transverse energy greater than 11\,GeV, 
\item{} EMC energy in the barrel calorimeter greater than 3.5\,GeV, 
\item{} EMC energy in the rear calorimeter greater than 2\,GeV.
\end{itemize}
In addition, at least one CTD track 
coming from the $ep$ interaction region was required. 
Events with timing measured by the veto counters, inconsistent
with an $ep$ interaction, were removed.
 
In the second level trigger beam-gas 
events were rejected by exploiting
the excellent timing resolution of the calorimeter. 
Also events were rejected in which the vertex determined by the CTD was
not compatible with the nominal $ep$ interaction region.

In the third level trigger (TLT) the full event information was available. 
Calorimeter timing cuts were tightened for further 
rejection of the remaining beam-gas events. Events 
were required to have a transverse energy outside a cone of $\theta$\,$=$\,$10^{\circ}$
with respect to  the 
proton direction, \ETCONE, above 12\,GeV.
The online value overestimates the offline \ETCONE\
due to the simplified event reconstruction at the TLT.
In addition, the following requirements were made:
\begin{itemize}
\item{} $p_{\perp}^{max+} > 0.5$\,GeV,
\item{} $p_{\perp}^{max-} > 0.5$\,GeV and
\item{} $p_{\perp}^{max+} + p_{\perp}^{max-} > 2.0$\,GeV,
\end{itemize}
where the quantity $p_{\perp}^{max+}$ ($p_{\perp}^{max-}$) 
is defined as the maximum transverse momentum of any positive (negative) track 
associated with the reconstructed event vertex  
in the polar angle range $15^\circ < \theta < 165^\circ$.

\section{Data Analysis}
 
\subsection{Offline Data Selection}
 
The event sample accepted by the 
trigger algorithm was processed using the 
standard offline ZEUS detector calibration and event reconstruction code.
To define an inclusive photoproduction sample, 
the following requirements were imposed:
\begin{itemize}
\item{} A reconstructed vertex with at least three tracks associated to it.
\item{} \ETCONE\ $ > 12$\,GeV.
%
\item{} No scattered positron found in the CAL
according to the algorithm described in ref.~\cite{DIRECT}.
This requirement removes deep inelastic scattering (DIS) neutral current       
events, thereby restricting \qq~to below 4\,GeV$^2$.
The corresponding median \qq~is about~$5{\cdot}10^{-4}$\,\g2.
\item{} $0.1 < $ \YJB\ $ < 0.7$. Here \YJB\
is the Jacquet-Blondel~\cite{JB} estimate of $y$,          the fraction 
of the positron energy carried by the photon in the proton rest frame.
It is defined as:
 
$$ y_{\mathsf{JB}} = {\Sigma_{i}(E-p_{z})_{i} \over 2E_{e}}, $$
 
\noindent
where the sum runs over all calorimeter cells
and $p_{z}$ is the $Z$ component 
of the momentum vector assigned to each cell of energy $E$. 
The lower \YJB\ cut rejects events from a region where
the acceptance is small because of the trigger 
requirements. The upper cut rejects possible background from DIS events in which the 
scattered positron has not been identified, and therefore is included in               
the \YJB\ calculation, thus producing a value of \YJB\ closer to~1.

\end{itemize}
With these requirements, an inclusive photoproduction 
sample of about 450,000 events was selected.
The \gp \ centre of mass 
energy of these events was calculated from \YJB\ via the expression 
$W_{\mathsf{JB}} = \sqrt{4y_{\mathsf{JB}}E_{p}E_{e}} $ 
and ranges from 100 to 250\,GeV.
A systematic shift observed in the reconstructed values of
$W_{\mathsf{JB}}$ with respect to the true $W$ of the event, 
due to energy losses in inactive material in front of 
the calorimeter and particles lost 
in the rear beam pipe, was corrected using Monte Carlo (MC)           
techniques~\cite{paper_93,DIRECT}. 
The centre of mass energy range covered by the 
photoproduction sample is then \wrang.

A $\ds$ reconstruction algorithm was applied to all selected events.
This algorithm combines the reconstructed tracks in each event to form $\ds$
candidates assuming the decay channels~(\ref{kpi}) or (\ref{k3pi}). It uses the 
mass difference technique to suppress the high background due to 
random combinations from non-$c\overline{c}$ 
events, which have   a much higher cross section.
Only tracks associated with the event vertex and
having $p_\perp > 0.2$\,GeV and 
$|\eta| < 1.75 $ are included in the combinations.

\subsection{ \bf \boldmath 
The ${\ds}^{+} \rightarrow \d0 \pi_{S}^{+} \rightarrow 
( K^{-} \pi^{+} ) \pi_{S}^{+}$ Decay Channel }
 
In each event tracks with opposite charges and $p_\perp > 0.5$\,GeV were 
combined into pairs to form $\d0$ candidates. 
The invariant mass $M(K\pi)$ of each pair
was calculated. No particle identification was used, 
so kaon and pion masses were assumed 
in turn for each particle  in the pair. 
A third track, assumed to be the soft pion, $\pi_{S}$, 
with a charge opposite to that of the particle taken as a kaon, was then added to 
the combination. The mass difference $\Delta M = M(K\pi\pi_{S}) - M(K\pi)$ was 
evaluated. Only $K\pi\pi_{S}$ combinations with              
$-1.5 < \eta < 1.0$, for which the acceptance is high, were kept.
As a result of the \linebreak
\ETCONE\ $ > 12$\,GeV cut                              
the acceptance of the $K\pi\pi_{S}$ combinations having $p_\perp$ 
below 3\,GeV is very small.
Thus $p_\perp (K\pi\pi_{S})> 3$\,GeV
was required.                                   
This cut excludes almost all soft pions with
$p_\perp (\pi_{S}) < 0.2$\,GeV, so the overall cut of
$p_\perp > 0.2$\,GeV on all tracks causes essentially
no loss of $D^*$ candidates.

In Fig.\,1a the $\Delta M$ distribution for combinations with 
$1.80 < M(K\pi) < 1.92$\,GeV                                
is shown. A clear peak at the nominal value of                                      
$M(\ds) - M(\d0)$ \cite{PDG} is observed. 
To determine the combinatorial background under the peak, 
combinations in the range $1.68 < M(K\pi) < 2.04$\,GeV were used, in which 
both tracks forming the $\d0$ candidates had the same 
charge (these will be referred to as                         
wrong charge combinations).                                             
The $\Delta M$ distribution from these combinations, normalized to 
the number of right charge combinations in the range                             
 $155 < \Delta M < 180$\,MeV, is also shown in Fig.\,1a.

The $M(K\pi)$ spectrum corresponding to combinations having a mass difference in the 
range $143 < \Delta M < 148$\,MeV  is shown in Fig.\,1b.        
A clear peak at the nominal $\d0$ mass of 1.8645~$\pm$~0.0005\,GeV~\cite{PDG}
is observed.  
The combinatorial background was again determined by using 
wrong charge combinations. The $M(K\pi)$ distribution from these    
combinations was normalized to the number of right charge combinations 
in the range $2.0 < M(K\pi) < 2.5$\,GeV. 
A clear excess of right charge combinations 
with respect to the combinatorial background           
was observed for masses below that 
of the $\d0$ meson. This excess is reproduced by the 
MC simulation (when a sample including 
all decay modes of the $D^0$ is used), 
and is mainly due to events in which a $\d0$ decays 
into $K\pi\pi^{0}$ and the extra $\pi^{0}$ is not included in the 
$\d0$ invariant mass reconstruction.
 
      The signals from Fig.\,1 were fitted with the maximum       
likelihood method                          to a sum of a Gaussian 
(describing the signal) and a functional form (describing the background
shape) of $A\cdot (\Delta M-m_{\pi})^{B}$ for Fig.\,1a and  
$\exp(C+D\cdot M(K\pi ))$ for Fig.\,1b.                                        
  The mass and width values obtained  
were: $\Delta M = 145.44\pm 0.08$\,MeV, $\sigma = 0.68\pm 0.08$\,MeV from
Fig.\,1a and $M(D^o) = 1860\pm 4$ MeV, $\sigma = 29\pm 4$ MeV from Fig.\,1b.
The mass values obtained are in agreement with the PDG values~\cite{PDG}.
The width values agree with our MC simulations.
The observed signals confirm that the reconstructed $\ds$ candidates come from              
channel~(\ref{kpi}).            
The contribution of other $D^o$ decay modes to this signal is negligible as
estimated by MC studies.
The number of reconstructed $\ds$'s was determined by subtracting
the normalized background distributions described above 
from the right charge            distributions.
The use of a wider $M(K\pi )$ range for the wrong charge combinations
compared with the signal region
reduces the statistical error of the subtraction procedure.
After subtracting the background from the $\Delta M$ distribution of   
Fig.\,1a,                                                                
a signal of {\totala} reconstructed $\ds$'s was obtained, consistent with
the corresponding subtracted signal from the $M(K\pi )$ distribution.

\subsection{ \bf \boldmath 
The ${\ds}^{+} \rightarrow \d0 \pi_{S}^{+} 
\rightarrow ( K^{-} \pi^{+} \pi^{+} \pi^{-} ) \pi_{S}^{+} $ Decay Channel }
 
For this channel, four       tracks were 
combined to form a $\d0$ candidate. 
The invariant mass $M(K\pi\pi\pi)$ was calculated 
for combinations having a total charge of zero.
Since no particle identification was used,
kaon and pion masses were assumed in turn 
for each particle  in the combination. 
A fifth track, assumed to be the soft pion, 
$\pi_{S}$,        with a charge opposite
to that of the kaon  was added to the combination and the mass difference 
$\Delta M = M(K\pi\pi\pi\pi_{S}) - M(K\pi\pi\pi)$ was determined. 
                Minimum transverse momenta  of
0.5\,GeV for the track taken to be the kaon and of 0.2\,GeV for the
other tracks were required.

Due to the higher 
number of decay particles in this channel, the combinatorial background is 
higher than that of channel~(\ref{kpi}).  
In order to achieve an improved signal to background ratio, 
we required the transverse momentum of the $K\pi\pi\pi\pi_{S}$ combination 
to be above 4\,GeV, and to satisfy the condition 
$p_{\perp}^{K\pi\pi\pi\pi_{S}} > 0.2 \, \cdot$\ETCONE. 
This latter requirement removed
about one third of the combinatorial background, 
i.e. the contribution from events with 
high \ETCONE\ in which a combination 
with relatively low transverse momentum was found. 
No reduction in the number of signal events was observed. 
Finally, the reconstructed $\ds$ candidates
were required to be in 
the same pseudorapidity range as for channel~(\ref{kpi}),
i.e. $-1.5 < \eta < 1.0$.
 
Fig.\,2a shows the $\Delta M$ distribution for those combinations with 
$1.81 < M(K\pi\pi\pi) < 1.91$\,GeV. The $M(K\pi\pi\pi)$ spectrum for 
combinations in the range $143 < \Delta M < 148$\,MeV is shown in Fig.\,2b.
A smaller window around the nominal $\d0$ mass was used in this channel 
because of the better mass resolution due to the lower average transverse 
momenta of the $\d0$ decay particles.
The signals from Fig.\,2 were fitted to a sum of a Gaussian 
(describing the signal) and a functional form (describing the background
shape) of 
\mbox{$A\cdot (\Delta M-m_{\pi})^{B}$} for Fig.\,2a and
$P_2 (x)=C+Dx+Ex^2$, where $x=M(K\pi\pi\pi)$ for Fig.\,2b. 
The mass and width values obtained  
were: $\Delta M = 145.42\pm 0.11$ MeV, \mbox{$\sigma = 0.83\pm 0.11$ MeV} from
Fig.\,2a and $M(D^o) = 1859\pm 3$ MeV, $\sigma = 19\pm 3$ MeV from Fig.\,2b.
The mass values obtained are in agreement with the PDG values~\cite{PDG},
and the widths agree with the MC values.                        
  
There are two main sources of background to the $D^*$ signals
in this decay channel.
The first one is
the combinatorial background coming from 
events or tracks in which no $\ds$ 
decaying through this channel is produced.
To determine this background,  wrong charge
 combinations with total charge  $\pm 2$ for the $\d0$ candidate and 
 total charge $\pm 1$ for the $D^*$ candidate were used.
The distributions of the wrong charge combinations were normalized to the number
of right charge combinations for $\Delta M$ in the range $155 < \Delta M < 180$\,MeV,
and for $M(K\pi\pi\pi)$ in the range $2.0 < M(K\pi\pi\pi) < 2.5$\,GeV and are shown
in Fig.\,2.
After the background subtraction, $199\pm 29$ reconstructed $D^*$'s
were observed in the  $\Delta M$ distribution.
The second source of background is due to events in which a $\ds$
decaying in this channel was produced,
but more than one combination was reconstructed 
inside the signal region, due to the erroneous assignment
of the kaon mass to a pion with the same charge from the $\d0$ decay.
 The fraction of $D^*$'s with a wrong ($K,\pi$) mass assignment was found
 from MC calculations to be $16\%$. After subtracting this contribution,         
 a final signal of $167\pm 25$ reconstructed $D^*$'s was obtained.
Monte Carlo studies show that the contribution of other $D^o$ decay modes to 
this signal is about $3\%$ and was neglected.

\section{ \bf Monte Carlo Simulation}
 
The Monte Carlo programs
PYTHIA~\cite{PYTHIA} and a recent version of HERWIG~\cite{HERWIG,giuanin}
containing an improved description of transverse energy distributions
were used to model 
the hadronic final states in $c\overline{c}$ production and to study
the efficiency of the data selection cuts.
Both programs are general purpose 
event generators which include     QCD LO 
matrix elements for charm photoproduction.
They          simulate higher order QCD radiation by 
 parton shower evolution  in the initial and final states,
taking into account coherence effects from the interference of soft gluon amplitudes.
Fragmentation into hadrons is simulated with a cluster 
algorithm~\cite{herclu} in the case of HERWIG, and with the 
LUND string model~\cite{LUND} in the case of PYTHIA. 
In HERWIG, the lepton-photon 
vertex is calculated exactly for the direct photon processes
and the equivalent 
photon approximation~\cite{EPA} is used for the 
resolved processes. In PYTHIA 
the Weizs\"{a}cker-Williams approximation~\cite{WWA} is used in both cases.

Large samples of $c\overline{c}$ events were generated
with both MC programs. Direct and resolved photon events, including
charm excitation,
were generated 
using several parton distribution parametrizations 
for both the proton and the photon.
The MRSG~\cite{MRSG} parametrization for the proton
and the GRV-G~HO~\cite{GRV}
for the photon were used to produce the reference samples.
These samples had at least 20 times the statistics of the data, 
so their contribution to the statistical error was neglected.
It was found from MC studies that, in the kinematic range used,
the results are insensitive to contributions from charm excitation in
the proton. However they are sensitive to charm excitation in the photon.
The differences between results
obtained with and without charm excitation in the photon are included
in the systematic errors. A large sample of $b\overline{b}$ events was also 
generated with PYTHIA to allow an estimation of the fraction of
photoproduced $D^*$'s originating from this process.
 
Events containing at least one charged $\ds$
decaying into     channel~(\ref{kpi}) or (\ref{k3pi}) were 
processed through the standard ZEUS detector and 
trigger simulation programs and through 
the event reconstruction package. 

\section{ \bf Cross Section Determination}
  
The integrated ${\ds}$ electroproduction 
cross section in the kinematic region 
defined by the selection cuts
\qqrang, \wrang, \etarang~and $p_{\perp}^{\ds} >$~3 or~4\,GeV~
is calculated using the formula:
$$ \sigma_{ep\rightarrow{\ds}X} = 
{ N_{corr}^{\ds} \over { {\cal L} \cdot Br } }, $$
 
\noindent
where $N_{corr}^{\ds}$ is the acceptance-corrected number of
$\ds$, $Br$ is the combined branching ratio of a given channel (see Table 1)  and
${\cal L} = 2.99 \pm 0.05$\,pb$^{-1}$ is the integrated luminosity. 

The usual method to correct data by the ratio of generated to 
reconstructed MC events is valid when the MC well describes the data
distributions of quantities used in the analysis. In our case
most of these distributions are well described by the MC events.
However, for the \ETCONE\ distribution the data yield
higher values compared to the events simulated with PYTHIA.
Therefore the reference MC used to calculate the acceptance
for channel~(1) was HERWIG.
In the case of channel~(\ref{k3pi}) PYTHIA was used
since HERWIG does not reproduce the decay width of the
resonances, and hence does not describe correctly the $D^0$ decays
into four particles.

In order to minimize the MC dependence of the corrections for the
\ETCONE\ cut in both channels, a two step procedure was used
to calculate $N_{corr}^{\ds}$:

\begin{itemize}
\item[(a)]{}                                                                                 
A weighting factor $\omega _1$ was obtained from the MC simulation  
and applied to each reconstructed $\ds$ candidate. These events  
were corrected
for the tracking efficiency of the $D^*$ reconstruction algorithm
and for all trigger and event selection cuts except for the \ETCONE\ cut.
The factor $\omega_1$ is defined as the number of generated $D^*$'s
divided by the number of reconstructed $\ds$'s. It is calculated in a
three-dimensional grid in the reconstructed quantities $p_{\perp}^{\ds}$,
$\eta^{\ds}$ and $W_{\mathsf{JB}}$. The average values of
$\omega_1$ are $2.99 \pm 0.03$ and $3.07 \pm 0.04$
for channel~(1) and channel~(2), respectively.
The variation of $\omega_1$ as a function of any of the three grid
variables is less than a factor of two in the restricted kinematic range.
 
\item[(b)]{}
To calculate the correction for the
\ETCONE\ $ > 12$\,GeV cut
we used an independent $D^*$ data sample selected for channel~(1) with no cut
on the energy deposition in the calorimeter at the TLT.
Only tracking information was used for this sample selection at the TLT.
The corresponding correction factor
$\omega_2$ is given by the total number of $D^*$'s (corrected as in (a) above) 
obtained with the TLT tracking selection divided by the number of those $D^*$'s
(corrected as in (a) above) in events which satisfy the cut
\ETCONE\ $ > 12$\,GeV.                             
The correction by this method increases the statistical 
error of the result, but reduces the      
 systematic error due to the different MC modeling 
of the \ETCONE\ distribution. The factor      
$\omega_2$ was applied as an overall weight for the total cross
section calculation or bin-by-bin in each differential cross section
distribution.
Since the $\omega_2$ correction is independent of the tracking 
acceptance (which is included in $\omega_1$), the $\omega_2$ weights
of channel~(1) were used also for the $N^{\ds}_{corr}$ evaluation of channel~(2).
The average values of $\omega_2$ are $1.8 \pm 0.2$
and  $1.4 \pm 0.1$ for $p_{\perp} > 3$\,GeV and  $p_{\perp} > 4$\,GeV,
respectively.                   
\end{itemize}

The reconstructed number of $\ds$ mesons 
after background subtraction
($N^{\ds}_{meas}$), the acceptance corrected 
number of produced $\ds$'s  
($N^{\ds}_{corr}=\omega_1\omega_2 N^{\ds}_{meas}$),                                       
the branching ratios $Br$~\cite{PDG} and the integrated cross sections for 
both decay channels are presented in Table\,1.                                              
For comparison with channel~(2), the value obtained for channel~(\ref{kpi}) 
requiring $p_\perp^{\ds} > 4$\,GeV is also shown.                  
The cross sections of both channels are in good agreement
for the same kinematic region.

The differential cross sections 
$d\sigma/dp_{\perp}^{D^*}$, $d\sigma/d\eta^{D^*}$ and $d\sigma/dW$                       
       were calculated using the same correction procedure in five bins
in $p_{\perp}^{D^*}$ (3-4; 4-5; 5-6; 6-8; 8-12 GeV), three bins in
$\eta^{D^*}$ ((-1.5)-(-0.6); (-0.6)-(0.1); (0.1-1.0)) and three bins in $W$
(115-170; 170-225; 225-280 GeV).                 
For each variable ($p_{\perp}^{D^*}$, $\eta^{D^*}$, $W$),
the other two variables are integrated over their
kinematic region.
The combinatorial 
background was subtracted bin-by-bin from each distribution using the method 
described in section~3.
The $d\sigma/dp_{\perp}^{D^*}$ distributions                                                 
             for both channels~(1) and (2) are shown in Fig.\,3.
The $d\sigma/d\eta^{D^*}$ and $d\sigma/dW$ distributions for
 channel~(1) are shown in Fig.\,4.
The results are in agreement with those of the
H1 untagged photoproduction data~\cite{H1_94},
also shown in Fig.\,3,
with the same $D^*$ rapidity range and with $95 < W < 268$ GeV, which
is quite similar to the kinematic range of this measurement.

%
%
 
\begin{table}[t]
\begin{center}
\begin{tabular}{|c|c|c|c|c|c|}
\hline
&&&&&\\
\noalign{\vskip-3.mm}
 Channel & $p_{\perp}^{\ds}$(GeV) & $N^{\ds}_{meas}$ & $N_{corr}^{\ds}$ & $Br$ &  $\sigma_{ep\rightarrow\ds X} $ (nb) \\
\noalign{\vskip-3.mm}
&&&&&\\
\hline                                                                                                  
\hline
&&&&&\\
\noalign{\vskip-3.mm}                                                                                   
 (\ref{kpi}) &$> 3$     & $152 \pm 16$ & $828\pm129$  & $0.0262 \pm 0.0010$ & $10.6\pm 1.7\pm ^{1.6}_{1.3}$ \\
&&&&&\\
\noalign{\vskip-2.mm}
(\ref{kpi})  &$> 4$     & ~\,$97 \pm 11$  & $348 \pm 53$~\, & $0.0262 \pm 0.0010$ & ~~~\,$4.5 \pm 0.7 \pm 0.6$    \\
&&&&&\\
\noalign{\vskip-2.mm}
 (\ref{k3pi}) &$> 4$     & \totalb  & $739\pm128$ & $0.051 \pm 0.003$& ~\,$4.8\pm 0.8\pm ^{1.0}_{0.6}$   \\
\noalign{\vskip-3.mm}
&&&&&\\
\hline
\end{tabular}
\end{center}
\caption{ Cross section  $\sigma_{ep\rightarrow\ds X}$ and related                 
 quantities for $Q^2 < 4$ GeV$^2$, $115 < W < 280$ GeV and $-1.5 < \eta^{D^*} < 1.0$ .}
\end{table}

\subsection{ Systematic Uncertainties} 
 
A detailed study of possible sources of 
systematic uncertainties was carried out, which we now
summarize.

Trigger acceptance uncertainties are mainly due to the 
different energy distributions predicted by the two event
generators PYTHIA and HERWIG
with and without flavour excitation.
 The uncertainties were determined
from the difference between the cross sections  obtained with both
MC generators.
For channel~(1) the largest shift was $-11\%$.
For channel~(2) the largest shift was $+10\%$.
 
To estimate the uncertainties in the tracking procedure, all track
selection cuts were varied. In the present analysis only tracks
associated with the event vertex were considered.
The systematic error due to any $D^*$ track not being fitted to the
vertex was estimated from a careful comparison of vertex fitting in
data and Monte Carlo.
The resulting uncertainties on the cross sections
are $^{+14}_{- 4}\%$ and 
$^{+18}_{-10}\%$ for channels~(1) and~(2), respectively.    
 
The uncertainty in the acceptance of the
\ETCONE\ $ > 12$ GeV cut was estimated by                        
changing the cut value by $\pm 0.5$\,GeV.
A variation in the cross section of         
 $^{+4}_{- 0}\%$ for channel~(1) and 
 $^{+3}_{-7}\%$                     
for channel~(2) was found. A shift of $\pm 3\%$ 
in the calorimeter energy scale produces a variation of          
 $^{+2}_{-0}\%$                     
        for channel~(1) and                                 
 $^{+0}_{-1}\%$                     
                                   for channel~(2).         
 
Background estimation uncertainties were determined by varying the
normalization region. The uncertainties were found to be             
 $^{+0}_{-1}\%$ for                 
both channels~(1) and (2).
 
The parton density parametrizations~\cite{PDFLIB}
MRS(G), MRS(A'),GRV94 HO and CTEQ3M for the proton and
GRV-G~HO, GRV-G~LO, LAC-G1, GS-G~HO and DG-G1 for the photon
were used in the MC event simulation. No significant variation of the 
acceptance was found in the kinematic region used for this analysis.                 
 
Finally, contributions of $\pm1.5 \%$ from
the luminosity measurement, and $\pm3.7 \%$ or
$\pm 5.7 \%$ from the branching ratios~\cite{PDG}
of channels~(1) or (2)   
respectively, were included.
 
All contributions to the systematic errors were
added in quadrature. The final systematic errors to the total cross sections   
are given in table 1. For 
the differential cross sections they were added in quadrature to the statistical
errors and are indicated as the outer error bars in Figures~3 and~4.

\section{ \bf Comparison with NLO QCD Calculations}

Full NLO computations of differential cross sections
for heavy quark production in any kinematic region~\cite{NLOdiff,kniehl}
became available recently.           
We compare our measurements with two such calculations.
 
One of the two computations
was done in the massive charm scheme~\cite{NLOdiff},
where $m_c$  acts as a cutoff for the
perturbative calculation. The program for $c\overline{c}$
photoproduction~\cite{NLO} from this
computation was used to produce total and differential $\ds$ cross sections in the
restricted kinematic region of our measurement.
From MC studies it was found that the contribution
of $b\overline{b}$ production to the $D^*$ cross section is below $5\%$. 
The charm fragmentation into $\ds$ was performed using the Peterson 
formula~\cite{PETER}:
 
$$ f(z) \propto { \left[ z \left( 1 - { 1 \over z } - { \epsilon_{c} \over ( 1 - z ) } \right)^{2} \right] }^{-1}, $$
                                 
\noindent
with $\epsilon_{c} = 0.06$~\cite{EPS} and using the branching ratio
$Br(c\rightarrow{\ds}^{+}) = 0.260\pm 0.021$
as measured by the OPAL collaboration~\cite{OPAL}.
 Here $z$ is the fraction of the charm quark
momentum taken by the $D^*$. In order to convert the photoproduction differential
cross sections into electroproduction cross sections, the $W$ range from
115 to 280\,GeV  was divided into  15\,GeV wide  bins.
The photoproduction cross sections were calculated in the center
of each bin and the results were added after multiplying by the proper
flux factors~\cite{paper_93}.         
 
A reference calculation was performed
with the MRSG and GRV-G~HO parton density parametrizations for
the proton and photon, respectively. 
The renormalization scale used was 
$\mu_{R}=m_{\perp}=\sqrt{ m_{c}^{2} + p_{\perp}^{2}} $ ($m_{c} = 1.5$\,GeV)
and the factorization scales of the photon and proton 
structure functions was $\mu_{F} = 2 \mu_{R}$.
The predicted reference cross sections 
thus obtained are compared to the measured ones in Figures~3 and~4.
Varying $m_c$ between 1.2--1.8\,GeV or $\mu_R$ between 0.5--2.0$m_\perp$
changes the cross sections in our kinematic range within $\pm 20\%$.
Decreasing $\epsilon_c$ to 0.035~\cite{OPAL} increases the cross section
by $15\%$.
Using different proton or photon
parton density parametrizations changes the cross section 
by less than $10\%$.
All of the cross sections predicted by \cite{NLOdiff} using the reference
parameters are lower than
the data and the cross section in the kinematic region of the  
$K\pi$ channel is about $50\%$ of the measured one. 
Better agreement with the data can be obtained with the choice
of the parameters
$\mu_{R}$ = 0.5\,$m_{\perp}$ and $m_{c} = 1.2$\,GeV
as shown in Figures~3 and~4.
With the present statistics the shapes of the NLO predictions
are in reasonable agreement with the data.

Recently, another type of NLO calculation~\cite{kniehl} was compared to 
the H1 and ZEUS preliminary results.
In this approach the charm
quark is treated according to the massless factorization scheme, which
assumes charm to be one of the active flavours inside the proton and
the photon, in contrast to the massive charm scheme.
%
%
%
The differential               
distributions obtained with the massless approach NLO calculation
are shown in Figures~3 and~4. 
The parton density parametrizations used for the proton and photon
%
%
were CTEQ4M~\cite{cteq} and GRV~HO~\cite{GRV}, respectively.
The renormalization and factorization scales, as well as the values of
$m_c$ and $\epsilon_c$ used are the same as in the                           
reference calculation of the massive-charm approach.
The agreement with our data is good. Using a different proton parametrization
(MRSG \cite{MRSG}) hardly changes the results~\cite{kniehl}
while the photon
parametrization ACFGP-mc~\cite{ACFGP} reduces
the cross section by $20\%$ in our kinematic range~\cite{kniehlpriv}.

\section{Summary}
 
The integrated and differential $\ds$ photoproduction 
cross sections in $ep$ collisions at HERA have been measured
with the ZEUS detector in the
restricted kinematic region $Q^2 < 4$\,GeV$^2$,
\wrang, \ptrang~and \etarang. The cross section 
$\sigma_{ep\,\rightarrow\,\ds\,X}$ obtained using the  channel
$\dskpi$ was measured to be 
\xsecs. Another $\ds$ decay channel, $\dsk3pi$,
has  been studied and good agreement with the $K \pi$ channel
has been found in the region of overlap ($p_{\perp}^{\ds} > 4$\,GeV).
A NLO perturbative QCD massive charm scheme calculation
predicts cross sections smaller than our measured values. 
Another NLO calculation
in which the charm quark is treated according to the massless
factorization scheme is in agreement with the data.
The shapes of the differential cross sections $d\sigma/dp_{\perp}^{\ds}$,     
$d\sigma/d\eta^{\ds}$ and $d\sigma/dW$ are reasonably
reproduced by both models.
                            
\section*{ \bf Acknowledgments }
 
The experiment was made possible by the inventiveness and the diligent efforts 
of the HERA machine group who continued to run HERA most efficiently during 1994. 
The design, construction and installation of the ZEUS detector has been made 
possible by the ingenuity and dedicated effort of many people from inside DESY 
and from the home institutes, who are not listed as authors. Their contributions 
are acknowledged with great appreciation.                                      
   We thank M.~Cacciari, S.~Frixione, B.A.~Kniehl and G.~Kramer for discussions and
S.~Frixione for providing us with the NLO code.  The strong support and encouragement 
of the DESY Directorate has been invaluable. We also gratefully acknowledge
the support of the DESY computing and network services.

\begin{figure}[tp]
\begin{center}
\vspace{-2.0cm}
\psfig{figure=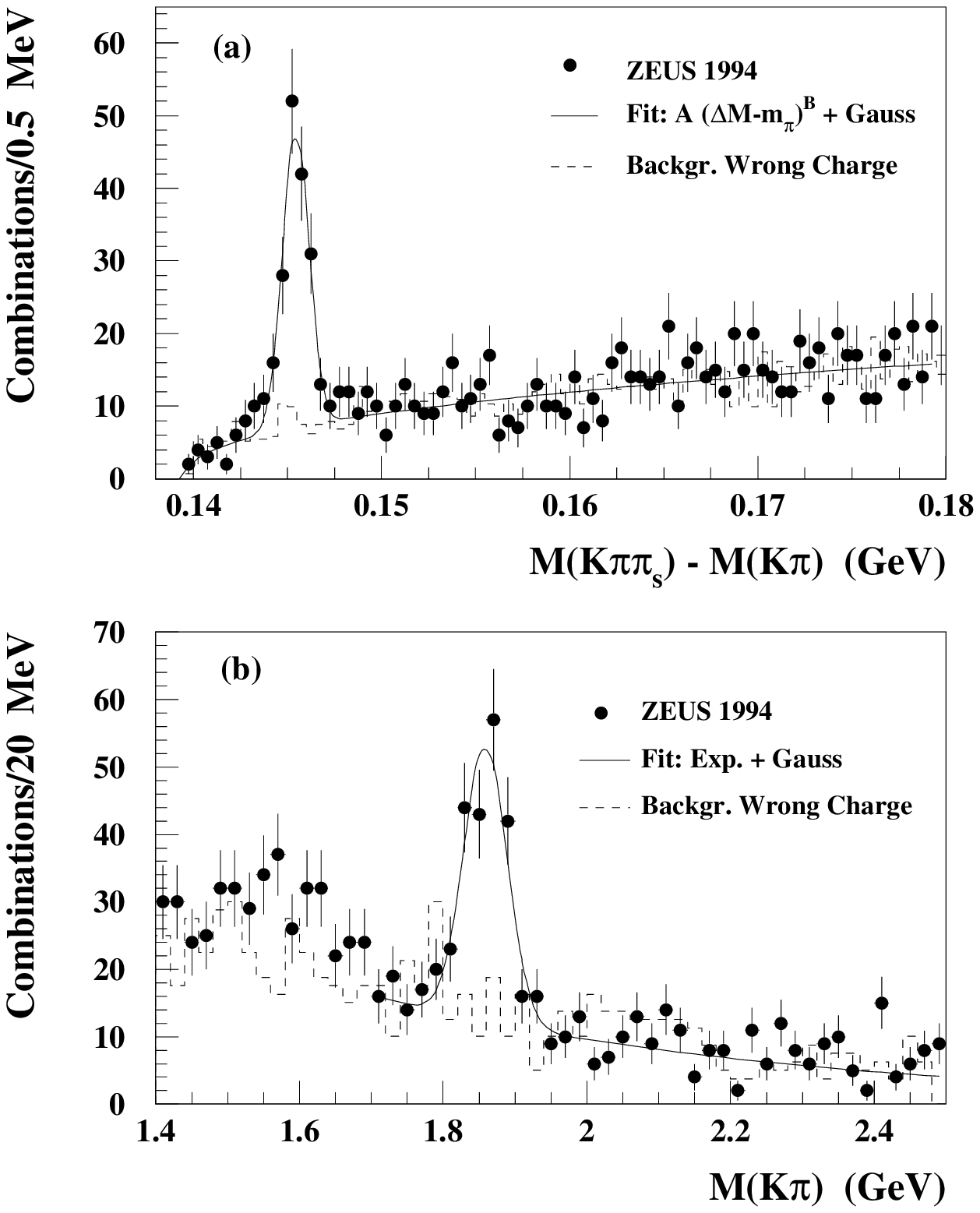 ,height=20cm}
\caption{ 
$\Delta M$ and $M(K\pi)$ distributions for channel~(1). 
In (a), the $\Delta M$ distributions for right charge combinations from the $M(K\pi)$ signal 
region (full circles)  and for wrong charge combinations from the region           
$1.68 < M(K\pi) < 2.04$\,GeV (dashed histogram) are shown.
In (b), the $M(K\pi)$ distributions for the right charge combinations from the $\Delta M$
signal region (full circles) and for wrong charge combinations from the $\Delta M$
signal region (dashed histogram) are shown.
The full lines are the results of fits to a sum of a Gaussian and the
functional form $A\cdot (\Delta M-m_{\pi})^{B}$ for (a) and
$\exp(C+D\cdot M(K\pi ))$ for (b).                                       
}
\label{kpi_signals}
\end{center}
\end{figure}

\begin{figure}[tp]
\begin{center}
\vspace{-2.0cm}
\psfig{figure=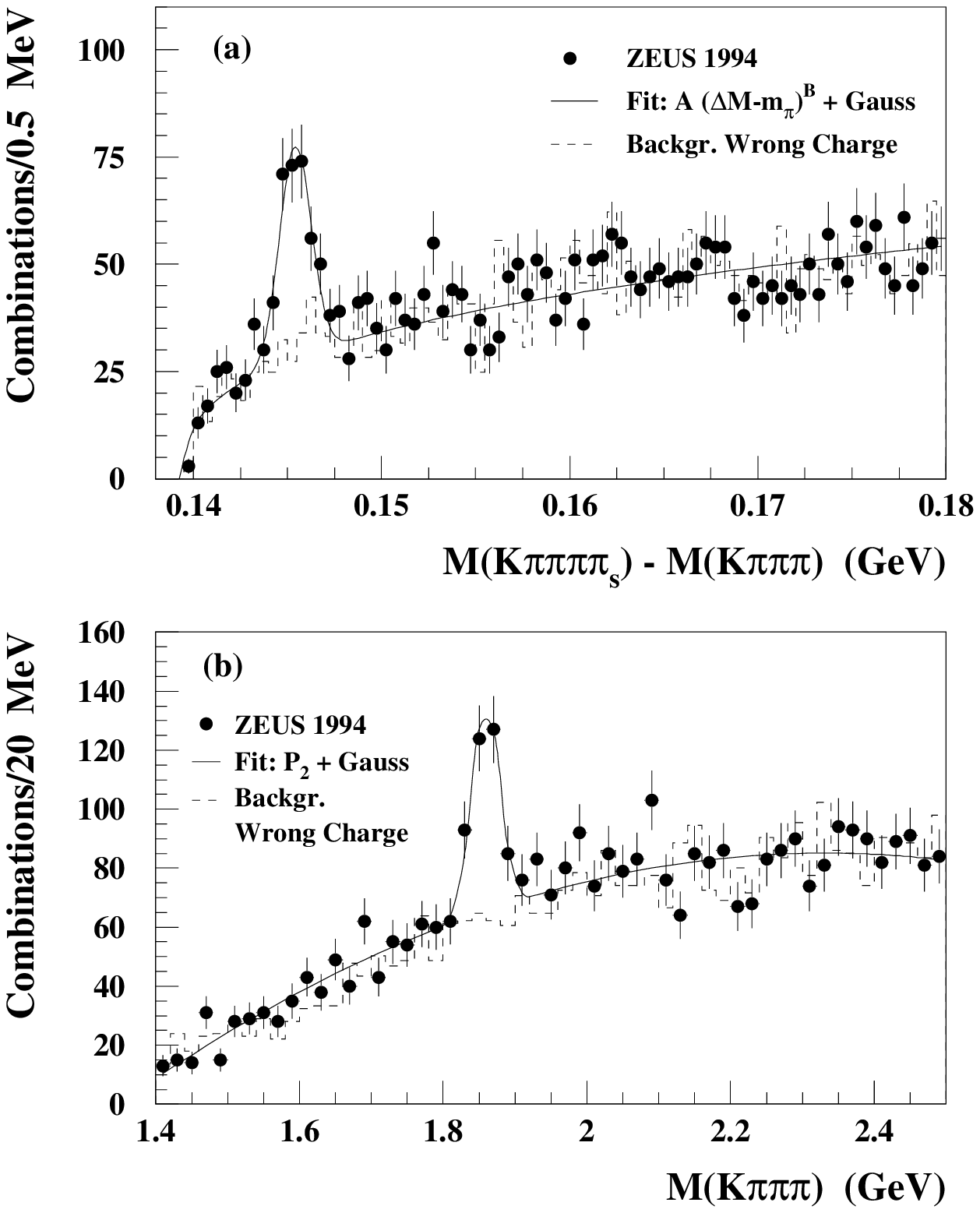 ,height=19.5cm}
\caption{ 
$\Delta M$ and $M(K\pi\pi\pi)$ distributions for channel~(2). 
In (a), the $\Delta M$ distributions are shown for the right charge combinations 
from the $M(K\pi\pi\pi)$ signal region (full circles), and for the background estimate 
using the wrong charge combinations (dashed histogram) as described in the text.
In (b), the $M(K\pi\pi\pi)$ distributions are shown for the right charge combinations 
from the $\Delta M$ signal region (full circles) and for the background estimate using 
the wrong charge combinations (dashed histogram) as described in the text.
The full lines are the results of fits to a sum of a Gaussian and the
 functional form                           
       $A\cdot (\Delta M-m_{\pi})^{B}$ for (a)      and
$P_2 (x)=C+Dx+Ex^2$, where $x=M(K\pi\pi\pi)$ for~(b).
}
\label{k3pi_signals}
\end{center}
\end{figure}

\begin{figure}[p]
\begin{center}
\vspace{-2.0cm}
\psfig{figure=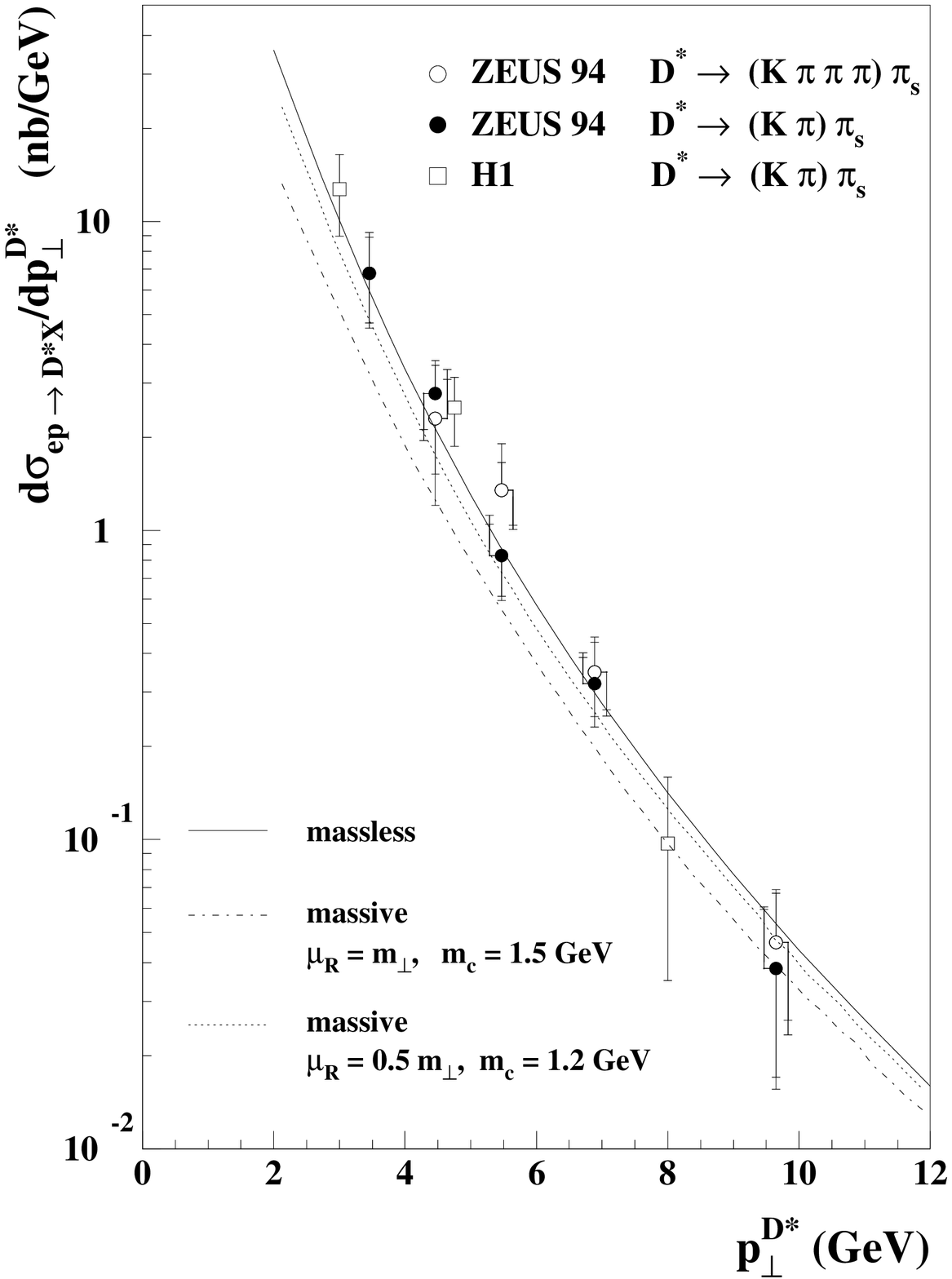 ,height=17cm}
\caption{ The differential cross section                                              
   $d\sigma_{ep \rightarrow D^* X}/dp_{\perp}^{\ds}$ 
                               of $\ds$ photoproduction, \qqrang, 
in the kinematic region                                                             
\wrang~and \etarang.                                
The experimental points are drawn at the positions of the average values
of an exponential fit in each bin.
The inner part of the vertical error bars shows the statistical error,
while the outer one shows the statistical and systematic errors added
in quadrature.
The H1 points~\cite{H1_94}
 include statistical errors and systematical errors due to trigger efficiency.
 The prediction of a NLO perturbative QCD calculation from a massive
charm approach~\cite{NLOdiff} is given by the dot-dashed curve,
                             using MRSG and GRV-G~HO as parton 
density parametrizations for the proton and photon respectively,     
fragmentation parameter $\epsilon_c =0.06$,
renormalization scale $\mu_R = m_{\perp}$ and $m_c =1.5$\,GeV.
The dotted curve is from the same calculation, but for
 $\mu_R = 0.5\,m_{\perp}$ and $m_c =1.2$\,GeV.
 The full curve comes from the massless
charm approach calculation~\cite{kniehl}, using the same parameters
as for the dot-dashed curve, but with
CTEQ4M taken for the parton density parametrization
for the proton. }
 
\end{center}
\end{figure}

\begin{figure}[p]
\begin{center}
\vspace{-2.0cm}
\psfig{figure=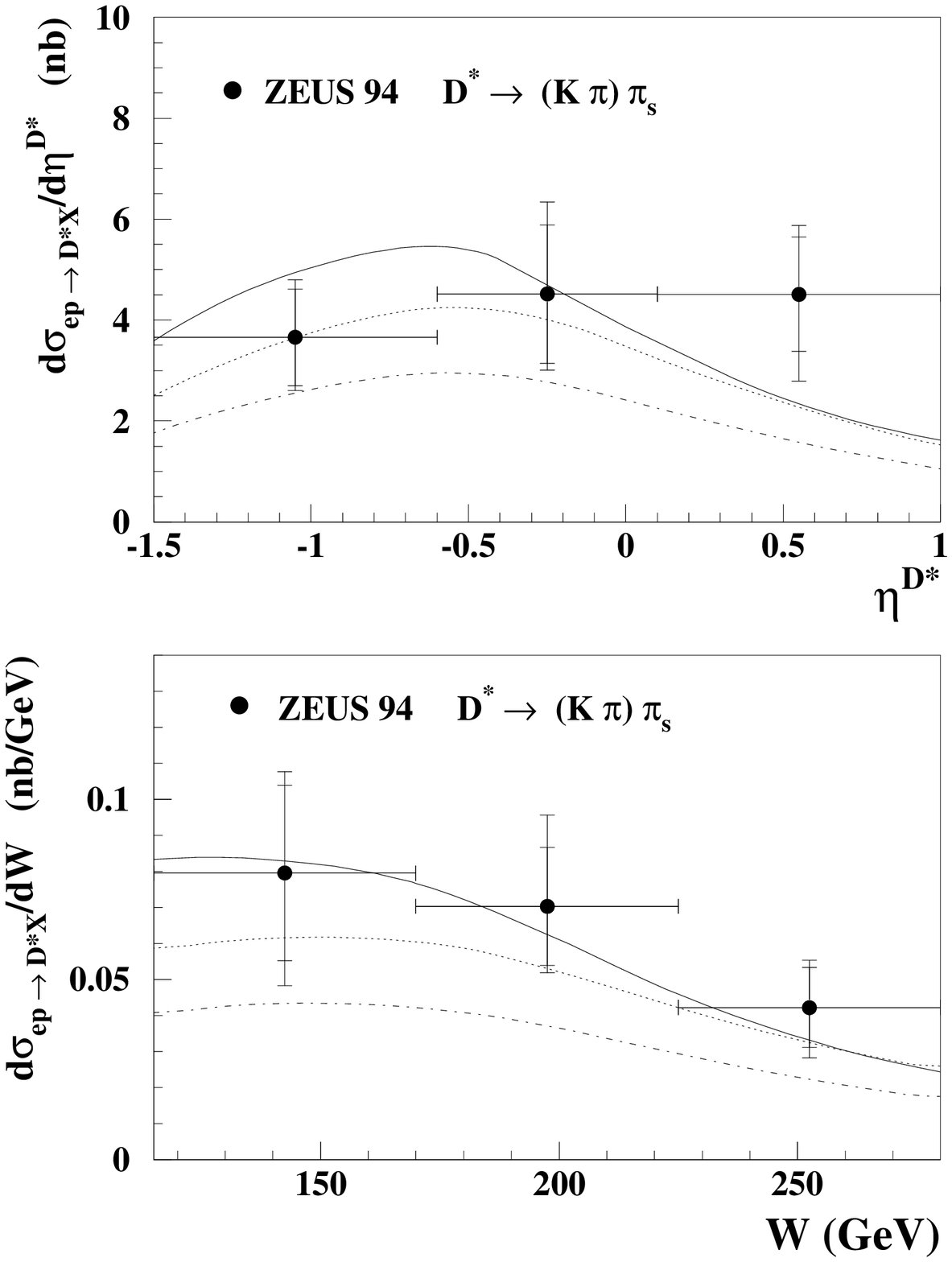 ,height=17cm}
\caption{ Differential cross sections                                             
for \qqrang~:                        
$d\sigma_{ep \rightarrow D^* X}/d\eta^{\ds}$
in the kinematic region \wrang~and \ptrang~(upper plot) and
$d\sigma_{ep \rightarrow D^* X}/dW$
 in the kinematic region \ptrang~and \etarang~(lower plot)    
for channel (1). The points are drawn at the centres of the corresponding bins.                                 
The inner part of the vertical error bars shows the statistical error,
while the outer one shows the statistical and systematic errors added
in quadrature. The prediction of a NLO perturbative QCD calculation
from a massive charm approach~\cite{NLOdiff} is given by the dot-dashed curve,
using MRSG and GRV-G~HO as parton density parametrizations for the proton
and photon respectively, fragmentation parameter $\epsilon_c =0.06$,
renormalization scale $\mu_R = m_{\perp}$ and $m_c =1.5$\,GeV.
The dotted curve is from the same calculation, but for
$\mu_R = 0.5\,m_{\perp}$ and $m_c =1.2$\,GeV. The full curve                     
comes  from the massless
charm approach calculation~\cite{kniehl}, using the same parameters
as for the dot-dashed curve, but with
 CTEQ4M taken for the parton density parametrization
for the proton. }
 
\end{center}
\end{figure}

\end{document}